\documentclass[%
 reprint,
superscriptaddress,
 amsmath,amssymb,
 aps,
prl,
]{revtex4-2}

\usepackage{epsfig,dsfont,amssymb,amsmath,amsthm,amsfonts,amsbsy,mathrsfs}
\usepackage{graphicx}
\usepackage{color}
\usepackage[dvipsnames]{xcolor}
\usepackage{bm}
\usepackage{multirow}
\usepackage{natbib}

\newcommand{\qq}{{\mathbf q}}

\newcommand{\rr}{{\mathbf r}}

\newcommand{\BEQ}{\begin{equation}}
\newcommand{\EEQ}{\end{equation}}
\newcommand{\BEA}{\begin{eqnarray}}
\newcommand{\EEA}{\end{eqnarray}}

\newcommand{\ee}{\mathbf{e}}
\newcommand{\vv}{\mathbf{v}}

\usepackage{tikz}
\usetikzlibrary{arrows,shapes,chains}

\begin{document}
\preprint{APS/123-QED}

\title{From flocking to glassiness in dense disordered polar Active Matter}
\author{Matteo Paoluzzi}
\email{matteo.paoluzzi@cnr.it}
\affiliation{Istituto per le Applicazioni del Calcolo del Consiglio Nazionale delle Ricerche, Via Pietro Castellino 111 80131 Napoli, Italy.}
\affiliation{Departament de Física de la Mat\`eria Condensada, Universitat de Barcelona, C. Martí Franqu\`es 1, 08028 Barcelona, Spain.}

\author{Demian Levis}

\affiliation{Departament de Física de la Mat\`eria Condensada, Universitat de Barcelona, C. Martí Franqu\`es 1, 08028 Barcelona, Spain.}
\affiliation{UBICS University of Barcelona Institute of Complex Systems, Mart\'i i Franqu\`es 1, E08028 Barcelona, Spain.}

\author{Ignacio Pagonabarraga}
\affiliation{Departament de Física de la Mat\`eria Condensada, Universitat de Barcelona, C. Martí Franqu\`es 1, 08028 Barcelona, Spain.}
\affiliation{UBICS University of Barcelona Institute of Complex Systems, Mart\'i i Franqu\`es 1, E08028 Barcelona, Spain.}

\date{\today}

\begin{abstract} %
\textbf{Abstract.} Living materials such as biological tissues or bacterial colonies are collections of heterogeneous entities of different sizes, capable of autonomous motion, and often capable of cooperating. Such a degree of complexity brings to collective motion on large scales. However, how the competition between geometrical frustration, autonomous motion, and the tendency to move cooperatively impact large-scale behavior remains an open question.  
We implement those three ingredients in a model of active matter and show that the system, in forming migratory patterns, can arrange in bands or develop long-range order, depending on the density of the system. We also show that the active material undergoes a reentrant glass transition triggered by the alignment interaction that typically causes only collective migratory motion. Finally, we observe that polar order destroys active phase separation, producing homogeneous, disordered moving configurations.
\end{abstract}

\maketitle

\textbf{Introduction.} Understanding the properties of collective rearrangements in dense active systems plays an important role in gaining insight into complex biological phenomena such as wound healing and metastasis invasion: two  paradigmatic examples where cell groups show 
coordinated motion to accomplish a specific task \cite{trepat2018mesoscale}.
The central role played by dynamical heterogeneities and glassy dynamics in collective cell motion has become progressively clear during the last decades ~\cite{angelini2010cell,Angelini11, park2015unjamming,malinverno2017endocytic}.
In contrast to equilibrium systems~\cite{Berthier2011}, dense living materials are collections of self-propelled objects whose dynamics is intrinsically out-of-equilibrium, as autonomous motion is based on biochemical processes that transform chemical energy into systematic motion \cite{Marchetti13,cates2012diffusive,Bechinger17,PhysRevX.12.010501}. 

These reaction processes do not satisfy detailed balance, at least on small scales where the single agent develops spontaneous motion. Moreover, living organisms are also characterized by alignment interactions with a purely mechanical origin \cite{Henkes2011}, or due to signaling mechanisms, as in the case of Vicsek interactions that capture collective animal behavior\cite{ballerini2008interaction,Vicsek95}. Although it is well established that the interplay between mechanical interactions, self-propulsion, and alignment interactions impact the structural and dynamical properties of active matter \cite{PhysRevX.6.021011,giavazzi2018flocking,PhysRevE.104.044606,park2015unjamming,barre2015motility,malinverno2017endocytic,martin2018collective,PhysRevE.104.054611,sese2018velocity,chate2008modeling,SFB15,BK13,shi2018self}, the generic features of emerging collective behaviors in dense active systems remain still poorly understood~\cite{henkes2020dense,PhysRevLett.124.078001,caprini2020hidden}.

Particulate models are very effective for modeling collective behavior since they require a limited number of parameters and provide a microscopic viewpoint on the key physical ingredients at play \cite{PhysRevLett.121.098003,PhysRevLett.125.178004,smeets2016emergent}. 
In this paper, we study collective behaviors emerging in a collection of geometrically frustrated active particles interacting via both, mechanical and aligning interactions. 
The former is purely repulsive through excluded volume.
As alignment interactions, following early works~\cite{resub1,resub2,resub3,PhysRevE.74.061908,Henkes2011}, we consider a simple feedback mechanism between particle velocity and self-propulsion direction
that tends to make the motion more persistent along the velocity direction.
These choices make the model minimal and thus give us access to general large-scale behaviors in active systems where these two interactions are at play. 
Moreover, the velocity/self-propulsion feedback introduced has proved to be very effective in reproducing large-scale behavior in confluent cell monolayers \cite{malinverno2017endocytic,giavazzi2018flocking,resub4}. Tests against experiments proved that collective cell behavior can be rationalized because of the minimal polar interaction we consider here \cite{malinverno2017endocytic}, even adopting a particle-based picture\cite{giavazzi2017giant}. 

As largely studied in the context of the celebrated Vicsek model, velocity-alignment interactions between point-like self-propelled particles, induce a phase transition towards a polarized, collectively moving state. Polar order has also been predicted in the case of implicit alignment interactions of the kind we consider here \cite{resub3}.
Although it is well established that band formation drives instabilities leading to a discontinuous transition in the case of Vicsek-like alignment interactions \cite{ginelli2016physics,PhysRevE.77.046113}, the phenomenology of polar active fluids becomes richer in the presence of birth-death processes \cite{PhysRevLett.108.088102}, disorder \cite{PhysRevLett.121.248002}, or additional conservation laws \cite{Chen16}.

In the presence of excluded volume interactions, self-propulsion triggers a phase separation, hence called Motility-Induced Phase Separation (MIPS), in the absence of velocity alignment \cite{cates2015motility}.  The mechanism behind MIPS is the self-blocking of particles when they collide, giving rise to a velocity that decays fast with the local density, eventually destabilizing the homogenous state. 
Alignment interactions favor close-by particles to move along the same direction and are thus responsible for a different aggregation mechanism (the one by which polar bands emerge).

In this work, besides showing that
dense disordered active materials develop a complex and rich phase diagram, 
where the competition between mechanical frustration and alignment destroys large-scale density heterogeneities and promotes disordered and homogeneous structures, we also document the existence of long-range order in 2D in the velocity field across the flocking transition. 

The long-range order is developed when the system is so dense that it behaves as an incompressible disordered active material. Using Finite-Size Scaling techniques, we compute the static critical exponents of the model showing that they satisfy scaling laws.

The transition qualitative changes from second to first-order as the density decreases below a threshold value.
Looking at the structural properties of the system across the transition, we show that for forming bands the local structure of the active liquid changes at the transition as signaled by a peak in the structure factor at low wave numbers. This peak disappears as the transition becomes of the second-order kind.
We thus make a connection between the structural properties of the system and the growth in collective polar order. We show that band formation, and thus a first-order-like scenario is signaled by a demixing transition. When the transition becomes second-order, demixing and bands disappear, and the fluid remains incompressible.

We also find that alignment plays a key role in the collective rearrangements of the system at high densities, in a regime where the active system develops glassy dynamics. As we show, alignment controls a reentrant glass transition: It first fluidizes the system at moderate coupling strengths but then triggers a glass transition towards a polarized amorphous solid for strong enough alignment. Finally, we documented how the alignment interaction considered here tends to destroy any kind of active phase separation driving the system toward homogeneous configurations.

\section*{Results}
The reason why we chose a polydisperse mixture is because we want to prevent crystallization in a wide range of the phase diagram. Because of the presence of geometrical frustration, even though the active fluid regime should maintain the same structural and dynamical properties as its monodisperse counterpart (as in the case of an equilibrium fluid in contact with a thermal bath where choosing a monodisperse system instead of a polydisperse one does not change its large scale behavior of the system), besides allowing a glass regime we prevent hexatic order in the phase separation (MIPS) region. \cite{paoluzzi2022motility}.

The model counts several control parameters. For exploring the collective behavior of the system we choose changing the packing fraction $\phi$, which is a natural control parameter in active systems that can be easily tuned in experiments, the strength of the alignment interaction $J$, which has been shown to play an important role in experiments of confluent monolayers \cite{malinverno2017endocytic}, and the persistence time of the active motion $\tau$, that sets also the distance from the equilibrium of the active system \cite{Fodor16}.

\subsection*{Continuous transition to a collective migratory phase at high-density}

\begin{figure*}[!h]
\centering\includegraphics[width=1.\textwidth]{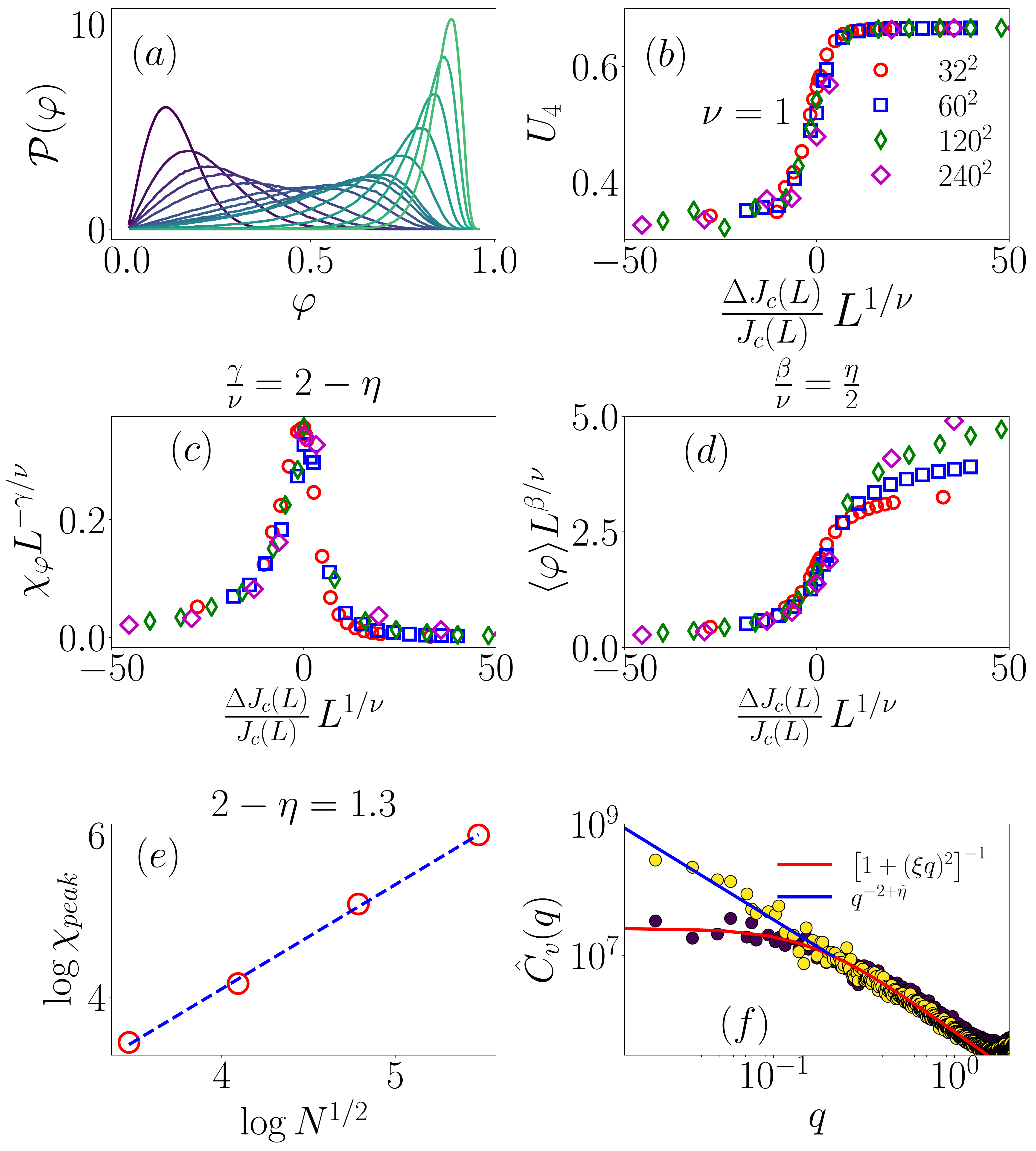}
\caption{ \textbf{Flocking transition at high density.} 
Here the packing fraction $\phi$ is $\phi=0.79$.
(a) The probability distribution function of the polar order parameter $\mathcal{P}(\varphi)$ changes continuously by crossing the transition (the alignment interaction strength changes within $J\in[0.1,1.0]$, from violet ($0.1$) to green ($1.0$)). 
(b) Finite-size scaling (FSS) of the Binder cumulant $U_4$ for different system sizes
(number of particles $N=32^2,...,240^2$, with box sizes $L=32,...,240$, respectively). 
(c) FSS of the susceptibility $\chi_\varphi$ which develops a peak at the transition. (d)  FSS of $\varphi$ for different system sizes sizes. 
(e) Scaling of the peak of $\chi_\varphi$ for increasing system sizes provides evidence for $2 - \eta \simeq 1.3$. 
(f) The Fourier transform of the spatial correlation function of velocity $\hat{C}_v(q)$ changes from Ornstein-Zernike
with finite correlation length $\xi$ (dashed red curve is fit to $\Delta J_c=-0.2$ data, violet symbols) to a power law at criticality ($\Delta J_c=0$, yellow symbols) that is fitted by $\tilde{\eta}-2=1.4\pm 0.1$. Data obtained for $L=240$.}
\label{fig:fig1}      
\end{figure*}

In the dense regime, $\phi\!=\!0.79$, the flocking transition can be characterized in terms of $J$ at fixed $\tau\!=\!1.0$, measuring the polar order through the polarization $\varphi(t)$, 
(in the following we indicate with $\varphi\equiv \varphi(t) $ and with $\langle \varphi \rangle$ its average performed over stationary trajectories. 
The location of the flocking transition is obtained from the  {\it magnetic} susceptibility peak, $\chi_\varphi \!\equiv\! N \langle \left[ \varphi - \langle \varphi \rangle \right]^2\rangle$. 
Fig.~\ref{fig:fig1}a reports the probability distribution function of the order parameter, $\mathcal{P}(\varphi)$, across the transition. $\mathcal{P}(\varphi)$ changes continuously as in an equilibrium second-order phase transition.  To make quantitative progress, we perform a finite-size scaling analysis, 
i.e., $N\!=\!32^2,60^2,120^2,240^2$,  changing the box size accordingly, i.e., $L\!=\!32,60,120,240$, at constant packing fraction, $\phi\!=\!0.79$, and compute  $\langle \varphi\rangle$, the Binder cumulant, $U_4\!=\!1-\langle \varphi^4 \rangle /(3 \langle \varphi^2 \rangle^2)$~\cite{binder1997applications}, 
and the susceptibility, $\chi_\varphi$ for each system size. As a proxy of the flocking transition, we consider the intersection of $U_4$ for different $N$ whose behavior provides the estimate of the critical point $J_c=0.75$ \cite{binder1997applications}. This estimate matches the position of the peak of $\chi_\varphi$. 

The continuous growth of  $\varphi$ with  $J$ is consistent with a  second-order phase transition.
To check its validity, we study the scaling of the peak of $\chi_\varphi$ with system size \cite{amit2005field}. 
Fig.~\ref{fig:fig1}e shows that the scaling follows a power-law, $\chi_{peak} \!\sim\! L^{2-\eta}$, with $2-\eta\!=\!1.3$.  
Employing the usual scaling {\it ansatz} that assumes for a generic observable $\mathcal{O}$ the finite-size scaling $\mathcal{O} \!=\! L^{x_\mathcal{O} / \nu} \hat{F}_\mathcal{O}(L^{1/\nu} (J- J_c))$ with $\hat{F}_\mathcal{O}$ a scaling function (we ignore sub-leading corrections and assume the divergence of the correlation length, $\xi\sim |J-J_c|^{-1/\nu}$, at the critical point~\cite{SM,amit2005field}) and the resulting scaling collapse $L^{-x_\mathcal{O}/\nu} \mathcal{O}$ {\it vs.} $L^{1/\nu} (J-J_c)$.
Fig. \ref{fig:fig1}b shows a good scaling collapse of $U_4$  with $\nu \simeq 1$. In the theory of critical phenomena, $\nu$ and $\eta$ are the only two independent critical exponents, while the others are bonded by scaling laws. For instance, from the scaling laws one has $\gamma / \nu = 2 -\eta$, with $\gamma$ the critical exponent of the susceptibility, and $\beta/\nu=(d+\eta -2)/2$, with $\beta$ the critical exponent of the order parameter ($d\!=\!2$ the spatial dimension). Figs.~\ref{fig:fig1}c-\ref{fig:fig1}d show the scaling laws are satisfied for $\chi_\varphi$ and the order parameter, respectively. 

The presence of a diverging correlation length is signaled by a qualitative change in the spatial correlation function of $\varphi$. 
Fig.~\ref{fig:fig1}f depicts the Fourier transform of the velocity spatial correlation function, $\hat{C}_v(q)$, ($q$ being the wave vector modulus~\cite{yllanes2017many}), for $L=240$. Far from the critical region, the correlation function has an  Ornstein-Zernike shape, $\hat{C}_v(q) \sim \left[ 1 + (\xi q)^2\right]^{-1}$, i.e., as in the Gaussian theory above the critical point, with $\xi$ being the correlation length.  
As the system approaches the critical point, $\hat{C}_v(q) \sim q^{-2+\tilde{\eta}}$  becomes power-law on large scales ($q\to0$), with the best fit $\tilde{\eta} = 1.4 \pm 0.1$,   compatible with the $\eta$ estimated independently from the scaling of $\chi_{\varphi}$.

Simulations at $\tau=1$, i.e., away from the Motility-Induced Phase Separation (MIPS) regime ~\cite{paoluzzi2022motility}, (the regime where active particles systems phase separate without microscopic attractive forces~\cite{cates2015motility} as the result of the competition between excluded volume effects and persistent motion), are run to analyze the role of density to promote collective motion.

\begin{figure*}[!h]
\centering\includegraphics[width=1.\textwidth]{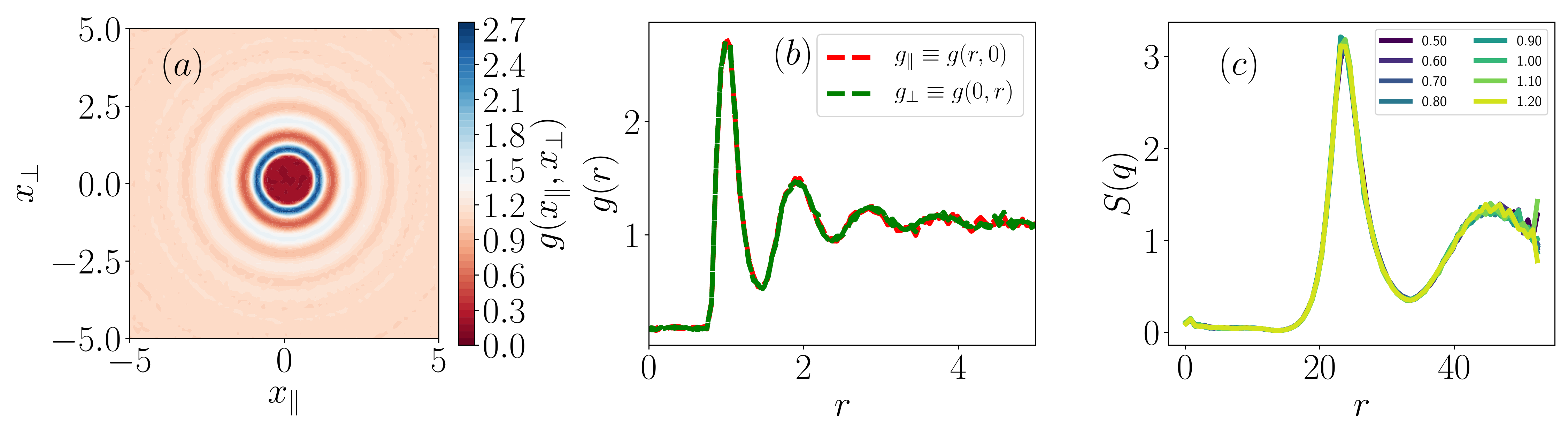}
\caption{ \textbf{Positional order of the migratory fluid.}
(a) Radial distribution function computed in the flocking frame (in panel (b) the components $g_\parallel$ and $g_\perp$) for alignment interaction strength $J=100$ (extremely deep in the flocking phase). (c) Static structure factor $S(q)$ across the transition (as in panels (a) and (b), we have persistence time $\tau=1$, number of particles $N=60^2$, and box size $L=60$). Increasing values of  $J$ from violet to yellow (see legend).
}
\label{fig:fig3_str}      
\end{figure*}

In Vicsek-like models, the emergence of polar order is also accompanied by band formation and anisotropic decay of the spatial correlation function. In the dense regime, once we measure the structural properties of the system within the migratory frame, we do not detect any anisotropy in the system, as shown in Fig. \ref{fig:fig3_str} where we report the radial distribution function and the static structure factor of the system deep in the flocking phase.

\subsection*{Crossover to the discontinuous transition mirrors positional order}

\begin{figure*}[!h]
\centering\includegraphics[width=1.\textwidth]{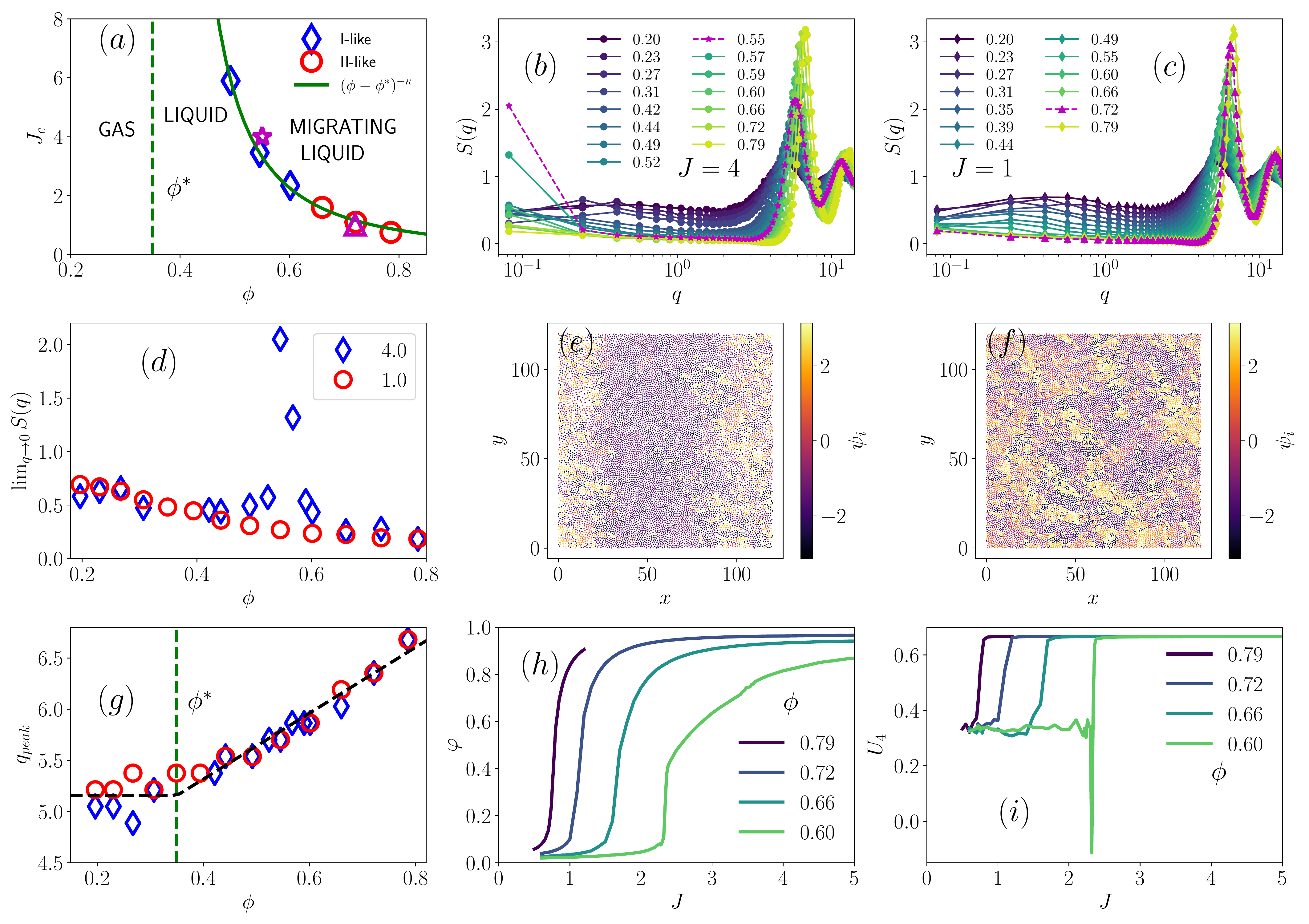}
\caption{ \textbf{Competition between positional and polar order.}
(a) Phase diagram in the alignment strength $J$ vs packing fraction $\phi$ plane (the box size is $L=120$, the persistence time $\tau=1$). 
From the study of the positional order and the center of mass dynamics, we can identify a Gas phase, a Liquid phase, and a migratory liquid state. Blue diamonds indicate a first-order-like flocking transition, and red circles a continuous, second-order-like, transition. The magenta star and triangle correspond to the magenta structure factor shown in panels (b) and (c), respectively. The solid green line is a fit to $(\phi - \phi^*)^{-\kappa}$, with $\kappa = 1.7$, and $\phi^*=0.35$.
(b) Static structure factor $S(q)$ crossing the first-order transition for $J=4$. The packing fraction increases from violet to yellow (see legend). The magenta curve indicates the value of $\phi$ where $S(q)$ develops a peak in the limit $q \to 0$. 
(c) $S(q)$ crossing the second-order transition ($J=1$). Increasing values of packing fraction from violet to yellow (see legend). At the flocking transition, $S(q)$ does not show any low-$q$ peak.
(d) Low-$q$ value of $S(q)$ as a function of $\phi$ for $J=1$ (red circles) and $J=4$ (blue diamonds). Crossing the first-order transition, $S(0)$ develops a clear peak.
(e)-(f) Snapshots of stationary configurations for $J=4$ (e) and $J=1$ (f) at the transition (the color map indicates the velocity orientation of each particle).
(g) Position of the peak of $S(q)$ as a function of $\phi$ for $J=1,4$.
(h) Vicsek order parameter as a function of $J$ for different values of $\phi$ (see legend) and the corresponding Binder parameter $U_4$ (panel (i)).}
\label{fig:fig3}      
\end{figure*}

We now study how changing the packing fraction of the system, and thus relaxing any incompressibility condition, changes the collective behavior of the active system. With this aim, we have performed simulations for spanning the $J$-$\phi$. In doing that, we monitor both, the positional order through the static structure factor, and the presence of migratory patterns through the Vicsek order parameter. The results of our analysis are shown in Fig. \ref{fig:fig3}. The phase diagram of the model is shown in Fig~\ref{fig:fig3}a.

We start our discussion with the flocking transition that brings the system from a resting liquid to a migrating liquid. We obtain the transition points in Fig~\ref{fig:fig3}a by monitoring the behavior of $\varphi$ and $U_4$ see Fig~\ref{fig:fig3}h and Fig~\ref{fig:fig3}i where we report the behavior of the two observables. We can thus discriminate between a continuous, second-order-like transition, where $\varphi$ is a smooth and growing function of $J$, and a discontinuous transition, i.e. first-order-like, where there is a jump in $\varphi$ that is mirrored by a negative value in $U_4$. We obtain that, by decreasing density,  the transition shifts to higher $J$ values, but it also changes from second to first order. In the following, we will specialize our study at $J=1$, where, as discussed in the previous section, the transition is second order, and $J=4$, where the transition becomes first order. Moreover, in spanning the $J$vs$\phi$ phase diagram, the system is always in a fluid phase (as we will see in the next section, there is a dynamical slowing down induced by the alignment interaction that takes place for larger $J$ values).

We thus monitor the positional order along the two non-equilibrium transitions. With this aim, we measured the static structure factor $S(q)$ for several values of $\phi$ and $J=1$ (where the transition is second-order-like) and $J=4$ (where the transition is first-order-like). The results are shown in Fig~\ref{fig:fig3}b,c. The behavior of $S(q)$ reveals that the qualitative difference of the two transitions is reflected by different structural properties. We highlight in magenta the $S(q)$ at the flocking transition.
In particular, the discontinuous transition ($J=4$) is accompanied by a peak in $S(0)$ that signals a demixing transition in the structure of the active fluid. A typical snapshot of the system in its stationary configuration confirms this behavior, as shown in Fig~\ref{fig:fig3}e where it is possible to appreciate the formation of a band. On the contrary, crossing the second-order-like transition ($J=1$), the system remains homogeneous, as shown in the snapshot in Fig~\ref{fig:fig3}f and reflected by the behavior of $S(q)$, Fig~\ref{fig:fig3}c.

The small wave number limit of $S(q)$ provides a measure of the compressibility of the system, its behavior is shown in Fig~\ref{fig:fig3}d.
As one can see, $S(0)$ is a monotonous decreasing function of $\phi$ for $J=1$, indicating that the system becomes progressively incompressible. Moreover, we can appreciate a  peak of the compressibility for $J=4$ that matches the flocking transition point.

As a generic property of the transition to the migrating fluid, we observe that the transition shifts to higher $J$  as $\phi$ decreases, following a  power-law, $J_c(\phi) \!\sim\! (\phi - \phi^*)^{-\kappa}$, with $\phi^* \simeq 0.35$ and $\kappa \!\simeq\! 1.7$.
The phenomenological scaling of $J_c(\phi)$ suggests that below a threshold value $\phi^*$, it is impossible to observe any flocking transition no matter how large the coupling constant of the alignment interaction is.
To check the existence of a finite threshold value $\phi^*$, we perform simulations for $\phi\!=\!0.44,0.39$ up to $J\!=\!30$, without detecting polar order. 
Additionally, we performed simulations at low density, i.e., $\phi=0.20$ up to $J=500$ without detecting any increase in polar order.

Similarly to what happens in cellular models \cite{giavazzi2018flocking}, the existence of an Active Gas phase where it is not possible to observe any migratory pattern follows directly by the definition of the model. In the gas regime, particles interact so rarely that we can approximate $\mu \mathbf{F}_i \simeq 0$, so that $\vv_i = v_0 \hat{\ee}_i$. This implies $\theta_i = \psi_i$ and thus the equation for angle $\theta_i$ becomes $\dot{\theta}_i = \eta_i$, the free rotational diffusion of a standard Active Brownian particle.

In other words, at packing fractions below $\phi^*$, the system is in the Active Gas phase where collisions between particles are so rare that self-alignment can not produce any polar order on length scales larger than single particle size. This is confirmed by the behavior of $S(q)$ that tends to become flat at low densities. 
For making quantitative this observation, we look at the position of the first peak of $S(q)$ (with $q \neq 0$), denoted $q_{peak}$. In the liquid phase, the peak is around $2 \pi$ and shifts to lower values in the gas phase. As shown in Fig~\ref{fig:fig3}g, $q_{peak}$ is almost density-independent at low packing fractions, i.e.,  below $\phi^*$. $q_{peak}$ starts growing at $\phi^*$ indicating that there is a qualitative change in the local structure of the active fluid around $\phi^*$.

\subsection*{Reentrant glass transition} 
In the previous section, we have shown that the emergence of collective behavior depends on the interplay between positional order and self-propelled (and self-aligned) motion. In particular, we observe a clear signature of the flocking transition looking at a pure positional order observable, as the static structure factor. In this section, we study how the self-alignment impact disordered configurations undergoing a glass transition.

\begin{figure*}[!h]
\centering\includegraphics[width=1.\textwidth]{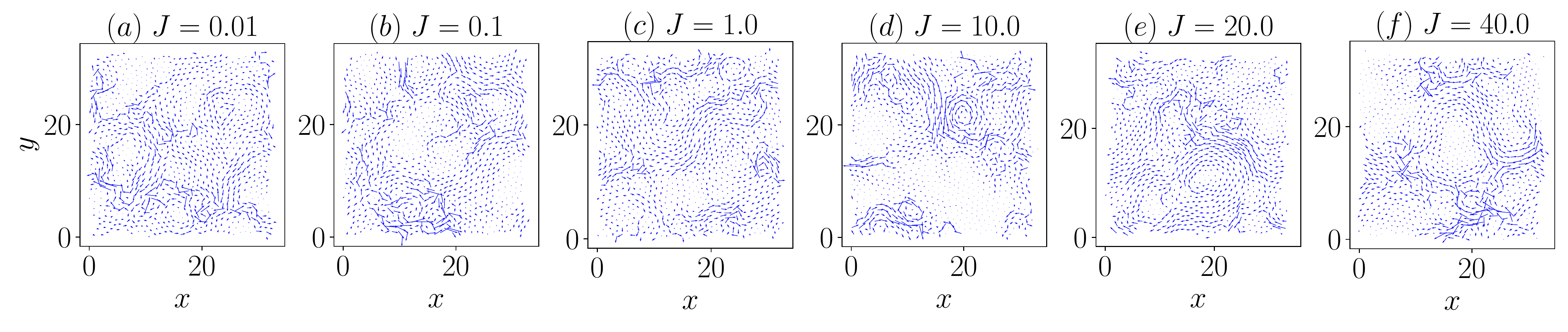} \\
\centering\includegraphics[width=1.\textwidth]{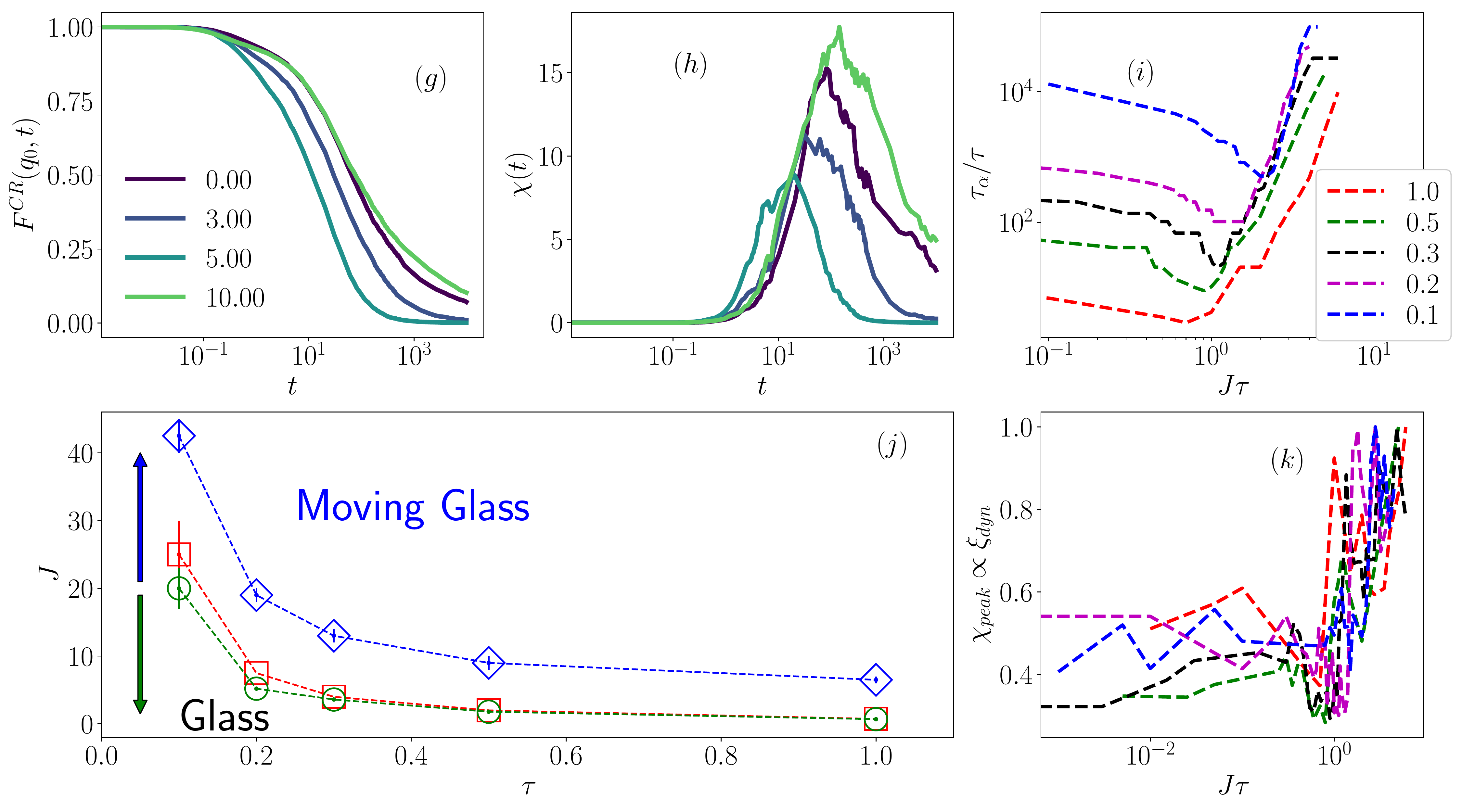}
\caption{ \textbf{Glass transition meets polar order.}
(a)-(f) Dynamical heterogeneity at persistence time $\tau=0.1$ for different values of alignment interaction strength $J$ in the center of mass frame.
(g) Intermediate scattering function (cage-relative) $F^{CR}$ computed at the first peak of the structure factor ($q_{peak} = 2 \pi \langle \sigma \rangle $, with $\sigma$ the particle size) for $\tau=0.1$ and packing fraction $\phi=0.79$ (the number of particle is $N=32^2$, we average over $N_s=126$ independent runs). Increasing values of $J$ go from violet to yellow (see legend). 
(h) Four-point dynamical susceptibility $\chi_4(t)$ for $\tau=0.1$ as an indicator of dynamical heterogeneity.
(i) Structural relaxation time $\tau_\alpha$ as a function of $J$ for different values of $\tau$ (see legend).
(j) Phase diagram using $\tau$ and $J$ as control parameters for $\phi=0.79$. The green symbols indicate the region where $\tau_\alpha$ reaches its minimum value. Red symbols indicate the flocking transition. Blue symbols indicate the glass transition defined through $\tau_\alpha$. Error bars reflect the finite number of $J$ sampled for computing the phase diagram.
The arrows indicate the direction of the dynamical slowing down. 
(k) The magnitude of the peak of $\chi_4$ as a function of $J$ (normalized to its maximum value for clarity, different curves indicate different values of $\tau$, 
as indicated in the legend of panel (i)).}
\label{fig:fig4}      
\end{figure*}

We now investigate the high-density region, $\phi\!=\!0.79$, by spanning the phase diagram 
through $J$ and $\tau$. As shown in Ref. \cite{paoluzzi2022motility} for $J=0$, 
the system behaves as a supercooled liquid at small $\tau$.
As it is shown in Fig. \ref{fig:fig4}a-f where we report the map of displacements, once we subtract the center of mass motion, the system develops dynamical heterogeneities that survive even for large $J$ values, i.e., in a region of the phase diagram characterized by collective migration.

To quantify the system's dynamical properties we compute the cage-relative intermediate scattering function $F^{CR}$ to remove the global motion due to the polar order \cite{vivek2017long,illing2017mermin} (details are provided in Methods).

Below the flocking transition, alignment fluidizes the system while, above the flocking transition, local rearrangements (reflected by a decay in $F^{CR}(q,t)$) are strongly inhibited as $J$ increases and the system eventually becomes a moving glass, i.e., $F^{CR}(q,t)$ develops a plateau for large $J$. 
The typical behavior of $F^{CR}(q,t)$ is shown in Fig. \ref{fig:fig4}g for $\tau\!=\!0.1$ and $\phi\!=\!0.79$. 
Fig. \ref{fig:fig4}i displays the corresponding relaxation time $\tau_\alpha$ (defined as $F^{CR}(\tau_\alpha)=e^{-1}$), where $\tau_\alpha$ is a non-monotonous function of $J$: starting from $J\!=\!0$, initially  $\tau_\alpha$ decreases as $J$ increases. 
The relaxation time reaches a minimum around the flocking transition and starts to increase again. This is because, in the polarized phase, particle rearrangements become strongly inhibited by collective motion \cite{giavazzi2018flocking,PhysRevE.104.044606}.

The existence of non-trivial glassy dynamics characterized by dynamical heterogeneity is confirmed by the study of the dynamical susceptibility $\chi_4(t)$, as shown in Fig. (\ref{fig:fig4})h (for $\tau=0.1$). $\chi_4(t)$ mirrors the non-monotonicity in $J$ observed in the behavior of the structural relaxation time $\tau_\alpha$. This behavior indicates a first stage, for $\tau J <1$, where the typical length scale $\xi_d$ of dynamical heterogeneity decreases with $J$ (the peak $\chi_4$ is proportional to $\xi_{dyn}$ \cite{Lacevic2003}) but then, around the flocking transition, i.e., $\tau J \simeq 1$, $\chi_4$ becomes broader, higher, and the peak shifts towards longer time. This fact indicates a proliferation of dynamical processes on different time scales (the width of $\chi_4$), larger dynamical correlation length (the height of the peak), and longer relaxation time (the position of the peak) \cite{Lacevic2003}.
We measure $\chi_{peak}$ for the entire data set, i.e., $\tau=0.1,0.2,0.3,0.5,1.0$, the result is shown in Fig. (\ref{fig:fig4})k where $\chi_{peak} \propto \xi_{dyn}$ has been normalized to the maximum value in each data set. As one can see, $\chi_{peak}$ undergoes a sharp crossover around the flocking transition, indicating that migratory patterns are characterized by larger cooperative regions.

These results are summarized in the resulting phase diagram, Fig.~\ref{fig:fig4}j. It is worth noting that a careful study of the structural properties of the system by means of the radial distribution function computed in the flocking frame does not provide any kind of anisotropy in the system, as shown in Fig. (\ref{fig:fig3_str}). We stress that the system undergoes a non-equilibrium dynamical slowing down towards a static glass phase for $J=0$ and $\tau \ll 1$, and it behaves as an active fluid for $\tau>1$ (as shown in Ref. \cite{paoluzzi2022motility}). As a result, from the point of view of the structural and dynamical properties of the system in the center of mass frame,  we obtain a reentrant phase diagram where, in the region $\tau<1$, the system results as a supercooled active fluid below the flocking transition line  $J<J_c(\tau)$, for $J_c(\tau)$ collective migration starts in a fluid phase that eventually becomes a moving glass for large enough $J$ values.

\subsection*{Suppression of active phase separation by polar order } 
We conclude 
by focusing on the impact of $J$ on phase separation.
Arguably one of the most studied collective phenomena exhibited by self-propelled particles is the so-called Motility-Induced Phase Separation (MIPS). Such a phenomenon hinders a self-trapping mechanism resulting in a fast reduction of the particles' velocity with increasing local density  \cite{cates2015motility}. 
The question of how alignment affects  MIPS has recently been addressed, showing that, well below the onset of flocking, polar alignment allows to trigger MIPS at smaller self-propulsion strengths \cite{sese2018velocity, PhysRevE.104.054611}. As we shall discuss below, strong enough coupling $J$ can also induce particle aggregation and phase separation via the local synchronization of particle motion, a different mechanism than the one controlling  MIPS \cite{martin2018collective}. Indeed, as alignment can induce strong velocity correlations that result in aggregates that move coherently, particles' velocity does not decay anymore with the local density, and thus the mechanism responsible for phase separation in the regime where the polarization is non-zero differs from MIPS \cite{sese2018velocity}.

We perform numerical simulations in a slab geometry, i.e., particles move in a rectangular box $L_x \!\times\! L_y$, with $L_x \!=\! 4 \, L_y$~\cite{SM}.
The rectangular box allows restricting the phase separation along one direction making it easy to individuate the surface between the two phases \cite{siebert2018critical,maggi2021universality,paliwal2017non}. 
We consider the case $L_y\!=\!60$, $N\!=\!100^2$ (corresponding to $\phi\!=\!0.54$), and $\tau=100$, i.e., deep in the phase separated region for $J\!=\!0$.

\begin{figure*}[!h]
\centering\includegraphics[width=1.\textwidth]{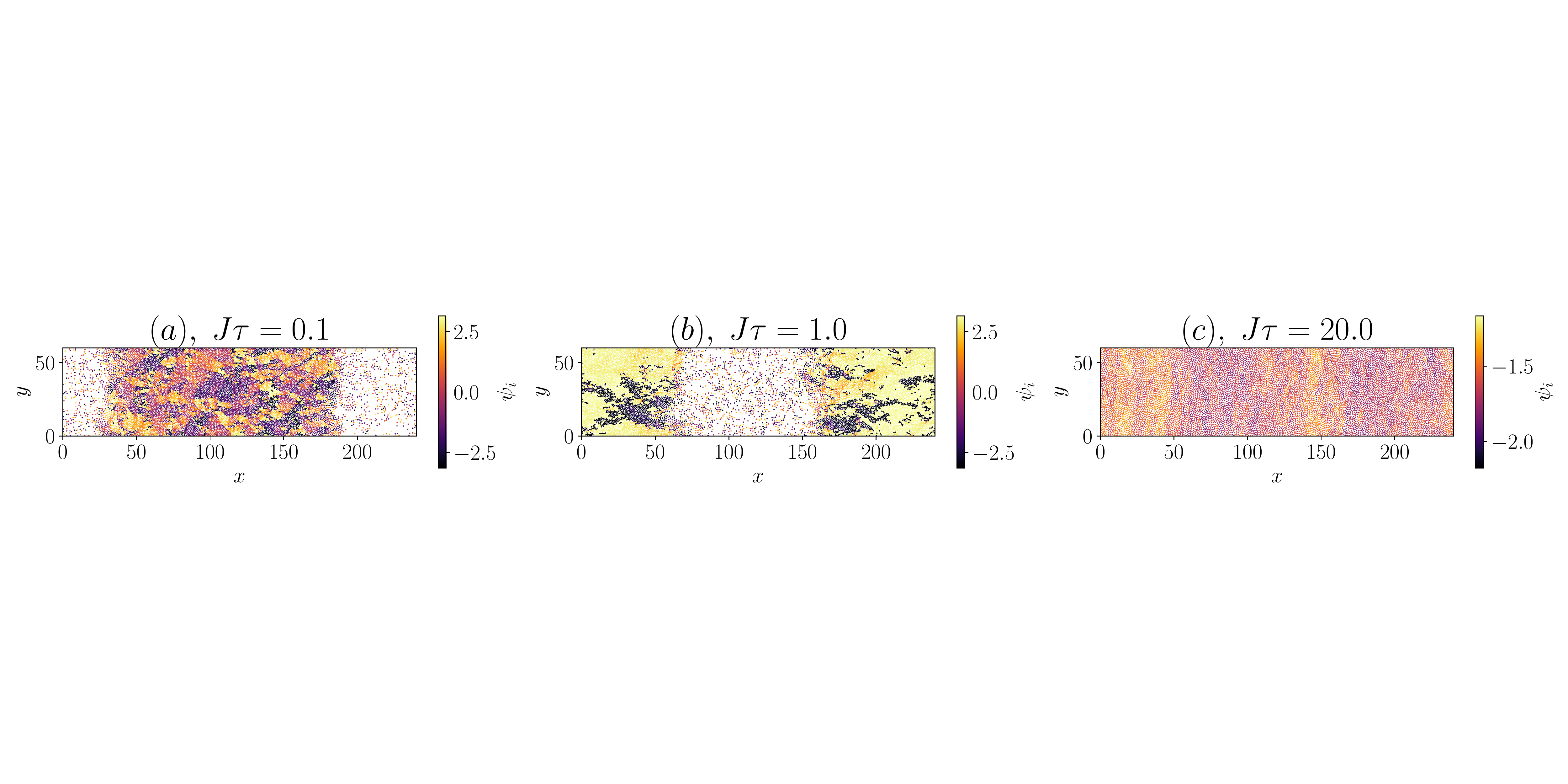}
\centering\includegraphics[width=1.\textwidth]{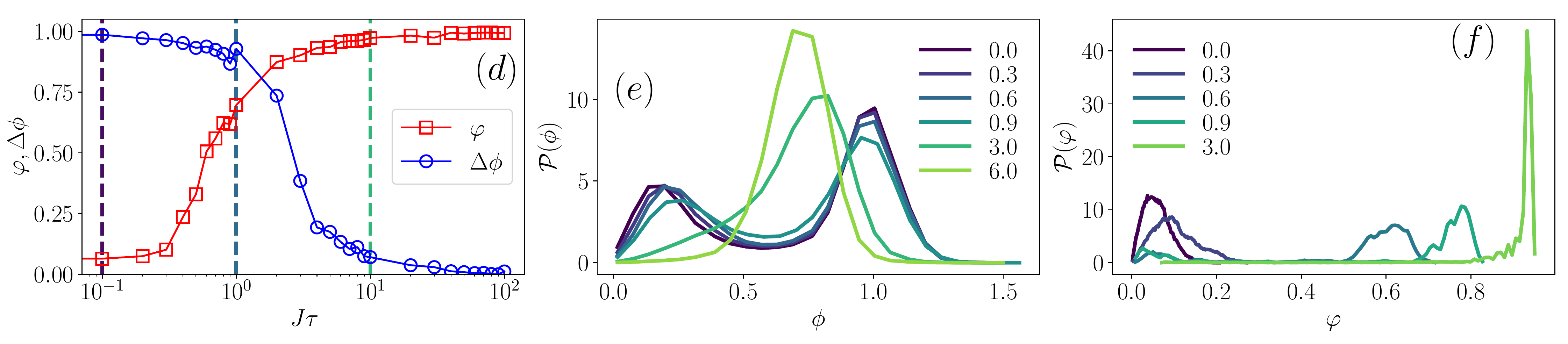}
\caption{ \textbf{Suppression of Motility-Induced Phase Separation in favor of homogeneous polar states.} 
 (a)-(c) Stationary configurations of the system in slab geometry (the packing fraction is $\phi=0.54$, number of particles $N=10^2$, box sides are $L_x=240$, $L_y=60$, and the persistence time is $\tau=100$) as the strength of the alignment interaction $J$ increases (from left to right, with a color scale indicating the orientation of the particles' velocity).
(d) Order parameter $\varphi$ (red squares) and distance between the two peaks $\Delta \phi$ of $\mathcal{P}(\phi)$ (blue circles) as a function of $J\tau$, at fixed $\tau=100$. Vertical dashed lines indicate $J \tau =0.1,1,10$ whose representative snapshots are shown in (a), (b), and (d), respectively. (e) Probability distribution function $\mathcal{P}(\phi)$ for increasing values of $J$ shows suppression of phase separation in favor of homogeneous configurations. (f) The corresponding distribution of the polar order parameter $\varphi$.}
\label{fig:fig5b}
\end{figure*}

In the presence of alignment interactions, the demixing order parameter $\Delta\phi$, i.e., the difference between the two peaks of the density distribution function, decreases as polar order increases. 
The phase separation present at moderate couplings is largely suppressed, see Fig 5a. Typical stationary configurations are shown in Fig.~\ref{fig:fig5b}a,b,c). 
For small values of $J$, i.e., $J=0.001$ that is far below the flocking transition, the system phase separates in a dilute and dense apolar phase, as in the standard MIPS (see Fig.~\ref{fig:fig5b}a). At larger $J$, we observe the formation of migratory polar patterns in the dense phase, as documented in Fig.~\ref{fig:fig5b}b (similarly to \cite{martin2018collective}). Finally, for $J \gg J_c$, phase separation disappears (see Fig.~\ref{fig:fig5b}c).
To provide a quantitative analysis of this observation, Fig.~\ref{fig:fig5b}e display the distribution $\mathcal{P}(\phi)$ for the local packing fraction, $\phi(x,y)\equiv \phi$ (in Fig. ~\ref{fig:fig5b}f the corresponding distribution $\mathcal{P}(\varphi)$). $\mathcal{P}(\phi)$ 
shows the typical features of a phase-separated state for small $J$ values where the distribution turns out to be double-peaked (see Fig.~\ref{fig:fig5b}e), while there is no hint of polar order (Fig.~\ref{fig:fig5b}e reports the corresponding distribution $\mathcal{P}(\varphi)$). The system is still phase-separated when the polar order rises (Fig.~\ref{fig:fig5b}b,d), although the nature of the dense phase changes, as it now displays polar order. Finally, $\mathcal{P}(\phi)$ becomes single peaked (Fig.~\ref{fig:fig5b}f), and thus the system is homogeneous from the point of view of the positional order, while polar order reaches its maximum intensity (Fig.~\ref{fig:fig5b}j).

This analysis is resumed in Fig. \ref{fig:fig5b}a where we show a comparison between the two relevant order parameters in this region of the phase diagram: the distance between the two peaks in the distribution of the local packing fraction $\Delta \phi$ 
, and the polarization $\varphi$. As we can see, as the system starts to develop a polar order, the distance between the two peaks starts to decrease and eventually phase separation disappears when the polar order saturates to its maximum value. We notice that there is a region of the phase diagram where phase separation and polar order coexist. This translates into the emergence of large-scale, dense migratory structures, in coexistence with a disordered background.
We stress that the existence of this coexistence region, where we can appreciate both, phase separation and collective migration, might suggest another mechanism of active phase separation driven by alignment interactions \cite{martin2018collective,PhysRevX.9.031043}. This aspect deserves future investigations.

\section*{Conclusion}
Despite disordered active materials are widespread in biology, suitable minimal models that allow for a systematic study of their collective properties through a few, essential, ingredients, remain poorly explored. 
Here we have established the structural and dynamical properties of a disordered 2D active system in the presence of two leading interactions: steric forces that prevent particles from overlapping, and alignment interactions. We observed that collective polar motion at high $\phi$ shows typical features of second-order phase transitions, i.e., the phenomenology is well captured by two independent critical exponents, and other quantities are bounded by scaling laws. 
We documented a continuous-to-discontinuous crossover tuned by density. We have shown that this crossover is intimately bound to the interplay between positional order and self-alignment interaction that can trigger band formation for a large value of the alignment strength and moderate packing fraction. Band formation is reflected by a peak in the structure factor that signals a demixing transition. Moreover, the presence of geometrical frustration destroys any anisotropy typical of the flocking transition. We also show that the presence of the alignment interaction allows us to distinguish between the Active Gas and Active Fluid phase since only in the latter is possible to appreciate the emergence of migratory patterns while in the former since the interaction is mediated by mechanical forces, there is no hope to induce collective polar order.

We stress that the existence of a second-order phase transition in dense and disordered active materials (with a diverging correlation length) might have important implications for collective rearrangements in biological systems, as in the case of confluent monolayers~\cite{malinverno2017endocytic}.
Recent hydrodynamic theories suggest that the presence of quenched disorder combined with the incompressibility condition leads to long-range order, and thus a second-order phase transition scenario, in two spatial dimensions \cite{chen2022hydrodynamic,PhysRevE.105.064605}. We stress that a minimal microscopic model like the one discussed here might be a suitable and natural benchmark for such a scenario.
This is because our model is basically incompressible at high packing fractions and geometrical frustration is a standard source of quenched disorder.

Consistently with the phenomenology of Vicsek models in two dimensions, we documented band formation as density decreases which leads to a first-order scenario for the flocking transition \cite{ginelli2016physics}. 
We stress that our finite-size scaling analysis does not suggest any crossover from second to first-order transition at high density, at least up to the system size investigated here.
As a future direction, it would be essential to provide an accurate estimate of the typical crossover length $L_c$, if there is any, leading to the first-to-second-order scenario at high density to understand how robust against finite-size fluctuations the phenomenology we documented is.

In the dense regime,
alignment interactions provide an additional control parameter for the glass transition in active matter. 
We showed that $\tau_\alpha$ is a non-monotonic function of $J$:
starting from a disordered active glass, alignment helps the system to fluidize while, for large enough $J$, $\tau_\alpha$ grows again. 
This non-monotonous behavior has been observed also in the proliferation of dynamical heterogeneities through the study of the dynamical susceptibility $\chi_4$.
Eventually, for large enough values of the alignment interaction, the system becomes a disordered moving solid. This fact shows how important alignment interactions are not only for developing collective motion but also for changing the structural properties of a living system \cite{PhysRevE.104.044606}. As a future direction, it might be interesting to study how the reentrant glass transition changes by changing the persistence length of the active motion $\ell = v_0 \tau$ by tuning $v_0$ instead of $\tau$.

Finally, we documented how alignment interactions strongly change structure formation in the system. This is possible because the feedback mechanism between velocity and self-propulsion promotes homogeneous configurations that tend to destroy phase separation. However, at least in the space of parameter explored here, we have also observed a region of the phase diagram where phase separation coexists with a polar state. As a future direction, it would be interesting to understand whether the presence of geometrical frustration plays a role away from the glassy regime. In particular, the tendency to promote homogeneous configurations should not depend on the presence of quenched disorder.

It might be also interesting in the future to perform the coarse-graining of the microscopic dynamics to get analytical insight into the model and compare it with other similar self-alignment interactions \cite{resub3}.

\section*{Methods}
In the following, we consider a model of active glass in two spatial dimensions where particles of different sizes interact via a purely repulsive interaction that prevents overlapping at any density. 
As an alignment interaction, we consider a simple feedback mechanism in the angular dynamics of the particle orientation. The alignment mechanism considered here has been introduced before in Refs. ~\cite{PhysRevE.74.061908,Henkes2011}. The alignment interaction makes the motion of each particle more persistent in the direction of the actual velocity. Since the direction of the actual velocity results from the sum of self-propulsion velocity and mechanical forces, the system can develop polar order as density increases.

\subsection{Microscopic Model}
The system is composed by a  polydisperse mixture \cite{PRX_Berthier} of $N$ Active Brownian disks, labelled  $i\!=\!1,...,N$, where particles interact via a purely repulsive power-law potential $v(r)$~\cite{Henkes2011,szamel2015glassy,flenner2016nonequilibrium,berthier2019glassy, janssen2019active,KumarPoly2021,paoluzzi2022motility}. The diameters $\sigma_i$, with average $\langle \sigma \rangle\!=\!1$, are extracted from a power-law distribution \cite{PRX_Berthier}. Particles move in a 2D square box of side $L$ with periodic boundary conditions. 
The dynamics of  disk $i$, individuated by the vector $\rr_i$ and moving with velocity $\vv_i$, is given by
\begin{align}
    \dot{\rr}_i & \!=\! \vv_i \!=\! v_0 \mathbf{e}_i + \mu \mathbf{F}_i \\ \nonumber 
    \dot{\theta}_i &\!=\! -J \sin(\theta_i - \psi_i) + \eta_i \; ,
\end{align}
where the angles $\theta_i$ and $\psi_i$ parameterize the self-propulsion direction $\mathbf{e}_i$ and the velocity direction $\vv_i / |\vv_i|$, respectively; $\eta_i$ is a white noise having zero-mean and $\langle \eta_i(t) \eta_j(s) \rangle \!=\! 2 \tau^{-1} \delta_{ij} \delta(t-s)$, with $\tau$ setting the persistence time of the active motion; 
 $J$ sets the strength of the alignment interaction that tends to make the motion more persistent along the velocity direction \cite{PhysRevE.74.061908,Henkes2011,giavazzi2018flocking}. 
The mechanical force is $\mathbf{F}_i\!=\!-\nabla_{\rr_i} \sum_{j \neq i} v(r_{ij})$ (with $r_{ij}\!=\!|\rr_i - \rr_j|$). In the following, we set $\mu\!=\!1$, $v_0\!=\!1$, 
and we adopt as control parameters $\tau$, $J$, and the  packing fraction $\phi\!=\!A_s N / A$, with $A_s=\pi (\langle \sigma \rangle / 2)^2$ the average particle's area \cite{paoluzzi2022motility}, and $A$ the simulation box area).

In order to prevent crystallization \cite{PRX_Berthier,wang2018low,khomenko2019depletion,PhysRevX.9.011049,PhysRevLett.122.255502,scalliet2019nature}, we consider a continuous mixture of disks of
different diameters $\sigma_i$ interacting through a pair potential 
\BEA
v(r_{ij}) &=& \left( \frac{\sigma_{ij}}{r_{ij}} \right)^n + G(r_{ij}) \\
G(r_{ij})    &=& c_0 + c_2 \left( \frac{r_{ij}}{\sigma_{ij}} \right)^2 +  c_4 \left( \frac{r_{ij}}{\sigma_{ij}} \right)^4
\EEA
with $r_{ij} \equiv | \rr_i - \rr_j|$. 
We set the softness exponent to $n=12$. 
The coefficients $c_0$, $c_2$, and $c_4$ are chosen in a way that $v(r_c)=v^\prime(r_c)=v^{\prime\prime}(r_c)=0$, 
($v^\prime(r) = \frac{d v}{dr} $).
To suppress the tendency to demix, we consider non-additive diameters  
\BEQ
\sigma_{ij} = \frac{1}{2} \left( \sigma_i + \sigma_j \right) \left[1 - \epsilon | \sigma_i - \sigma_j |  \right] \;, \,
\EEQ
where $\epsilon$ tunes the degree of non-additivity \cite{zhang2015beyond,PRX_Berthier}. 
The cutoff is $r_c=1.25 \sigma_{ij}$, and $\epsilon=0.2$. The particle diameters $\sigma_i$ are drawn from a power law distribution $P(\sigma)$
with $\langle \sigma \rangle \equiv \int_{\sigma_{min}}^{\sigma_{max}} d\sigma \, P(\sigma) \sigma = 1$, with $\sigma_{min}=0.73$,
$\sigma_{max}=1.62$, and $P(\sigma)= A \sigma^{-3}$, with $A$ a normalization constant \cite{PRX_Berthier}.

We explore the glassy regime using small system sizes at packing fraction $\phi=0.79$ (with $N=32^2$ and $L=32$) and performing averages over $N_s=10^2$ independent runs ($t_{max}=10^4 \Delta t$, with $\Delta t=2.5\times 10^{-3}$ the integration time step). We consider three persistence times, i.e., $\tau\!=\!0.1,0.3,0.5,1.0$ and varying $J$ inside the intervals
$J\!\in[\!0.01,20]$,  $J\in[0.01,35]$, $J\!\in\![0,20]$,
and 
$J\!\in\![0.01,100]$, respectively.

To investigate the nature of the flocking transition in dense and disordered configurations, we performed numerical simulations at high packing fractions, i.e., $\phi=0.79$, and $\tau=1$. In this region of the phase diagram, we have performed the Finite-Size Scaling analysis of the transition by changing the system sizes $L=32,60,120,240$ and thus $N=32^2,60^2,120^2,240^2$. 
We vary the magnitude of the alignment interaction within the interval $J\in[0.5,1.2]$.

For investigating the role of density on flocking transition, we performed simulations for $L=120$ and $N\!=\!95^2,100^2,105^2,110^2,115^2$ corresponding to $\phi=0.49,0.55,0.60,0.66,0.72,0.79$ varying $J$ at $\tau=1.0$.

The impact of alignment and the subsequent flocking transition in the active particle's aggregation and structure formation
has been investigated by performing numerical simulations in a rectangular box with periodic boundary conditions. In this slab geometry, the box is $L_x \times L_y$, with $L_x = 4 \times L_y$. We consider the case $L_y=60$, $N=100^2$ (corresponding to $\phi=0.54$), and $\tau=100$, i.e., deep in the phase separated region for $J=0$. 
In this simulations $J\in[0,10]$.

\subsection{Observables}
\subsubsection{Polar order}
For studying the polar order typical of the flocking state, the standard observable is the polar order parameter
\begin{align}
    \varphi(t) = \frac{1}{N} \left| \sum_{i} \frac{\boldsymbol{v}_i}{v_i} \right| \; .
\end{align}
For computing the polar order we consider stationary trajectories, i.e., we perform time-averages $\langle \varphi \rangle$ once $\varphi(t)$ reaches a stationary state.
To quantify the properties of the flocking transition we look at the probability distribution function $\mathcal{P}(\varphi)$ and its momenta. From the second and the fourth momentum we compute the Binder cumulant $U_4$ \cite{binder1997applications} defined as
\begin{align}
    U_4 = 1 - \frac{\langle \varphi^4 \rangle }{3 \langle \varphi^2\rangle^2} \; ,
\end{align}
that is scale-invariant at the critical point. 
We also measure the {\it magnetic} susceptibility $\chi_\varphi$ defined as
\begin{align}
    \chi_\varphi \equiv N \left\langle \left[ \varphi - \langle \varphi \rangle \right]^2 \right\rangle \; .
\end{align}
Since the model at high density reveals typical features of a second-order phase transition, we performed the usual Finite-Size Scaling analysis \cite{amit2005field}. Under the finite-size scaling {\it ansatz}, around the critical point, a given observable $\mathcal{O}$ for a system of linear size $L$ behaves as
\begin{align}
    \mathcal{O} = L^{x_\mathcal{O} / \nu} \left[ F_\mathcal{O}(L \xi^{-1}) + O(L^{L^{-\omega}},\xi^{-\omega})\right]
\end{align}
with $\xi$ the correlation length that diverges as $|J-J_c|^{-1/\nu}$ at the critical point. $F_\mathcal{O}$ is a universal finite-size scaling function, $x_\mathcal{O}$ is the critical exponent of the observable $\mathcal{O}$, and the exponent $\omega$ determines the sub-leading correction to the scaling. Once we ignore sub-leading corrections, and assume the divergence of the correlation length $\xi$, we get
\begin{align}
    \mathcal{O} = L^{x_\mathcal{O} / \nu} \hat{F}_\mathcal{O}(L^{1/\nu} (J- J_c)) \; , 
\end{align}
with $\hat{F}_\mathcal{O}$ another scaling function.
From the scaling {\it ansatz} follows the scaling collapse of different observables once plotted as $L^{-x_\mathcal{O}/\nu} \mathcal{O}$ vs $L^{1/\nu} (J-J_c)$.

To investigate the large-scale behavior of the system approaching the flocking transition we compute the Fourier transform of the velocity correlation function 
\begin{align}
    \hat{C}_v(q) = \left\langle \left| \sum_\rr e^{i \qq \cdot \rr} \vv(\rr) \right|^2 \right\rangle \; .
\end{align}
Where the velocity field $\vv(\rr) = (v_x(\rr), v_y(\rr))$ is obtained by discretizing the simulation box into a grid of linear size $2 \langle \sigma \rangle$. $\hat{C}_v(q)$ is thus obtained through the Fast Fourier Transform of $\vv(\rr)$ using data from the larger system size investigated, i.e., $L=240$.

\subsubsection{Moving Frame}
At a given time, we can define the polarization vector $\boldsymbol{\varphi}=(\varphi_x,\varphi_y)$ whose components are
\begin{align}
    \varphi_{x,y}(t) = \frac{1}{N}\sum_i \frac{v_i^{x,y}}{v_i} \; .
\end{align}
When the system develops polar order, we have also studied the dynamical and structural properties of the system in the flock frame. For a given observable $\mathcal{O}(\rr)$ that depends on a configuration of the system $\rr \equiv \{ \rr_i \}_{i=1}^N$, we can thus define $\mathcal{O}_{\parallel}$ and $\mathcal{O}_{\perp}$ that are computed longitudinally and perpendicularly to $\boldsymbol{\varphi}(t)$. The two components are computed in the following way.
\begin{itemize}
    \item At each time s$t$ we compute $\boldsymbol{\varphi}$ that, in two spatial dimensions, can be parametrized through the angle $\Theta$, i. e., $\boldsymbol{\varphi}(t)\equiv |\boldsymbol{\varphi}(t)| (\cos \Theta(t), \sin \Theta(t))$ (where the angle is taken in the lab frame).
    \item If $|\boldsymbol{\varphi}(t)| \neq 0$, we move to the frame of the polarization field. To do that, we
    perform a rotation of the reference frame where the new reference frame is rotated by $\Theta$.  Meaning that, a point originally located at $x,y$ in the lab frame, will have coordinates $(x_\parallel,x_\perp)$ in the flocking frame given by
    \begin{align}
        x_{\parallel} &= x \cos \Theta + y \sin \Theta\\ \nonumber 
        x_{\perp}     &= -x \sin \Theta + y \cos \Theta \; .     \end{align}
    \item In the case of a scalar observable, we can thus define $\mathcal{O}_\parallel \equiv \mathcal{O}(x_{\parallel},0,t)$ and $\mathcal{O}_\perp 
\equiv \mathcal{O}(0,x_{\perp},t)$.
\end{itemize}

\subsubsection{Phase separation}
Density heterogeneities are investigated by looking at the local packing fraction field $\phi(x,y)$ that is obtained by discretizing the simulation box in a lattice of size $\delta = 4 \langle \sigma \rangle$ and computing the local packing fraction in each box. We can thus define the probability distribution function $\mathcal{P}({\phi)}$ of the local packing fraction $\phi\equiv\phi(x,y)$, with the vector $(x,y)$ pointing at the center of each box, i.e, $(x,y)=\frac{1}{2}(i,j)\delta$, and $i=1,...,N_b$ (same for $j$), with $N_b = L / \delta$. In this way $\mathcal{P}({\phi}) \equiv \left\langle \delta \left[ {\phi} - \phi(x,y) \right] \right\rangle $.

\subsubsection{Dynamical transition}
For removing global motion due to the presence of polar order,
we consider only cage-relative (CR) quantities for computing dynamical observables \cite{vivek2017long,illing2017mermin}. These observables are computed considering the Voronoi neighbor of each particle so that no cutoff distances are required. The CR displacement $\Delta \rr_i^{CR}(t)$ for a given particle $i$ at time $t$ is defined as follows
\begin{align}
    \Delta \rr_i^{CR}(t) \equiv \rr_i(t) - \rr_i(t_0) - \frac{1}{N_i} \sum_{j=1}^{N_i} \left[ \rr_j(t) - \rr_j(t_0) \right] 
\end{align}
where we extend the sum over $j$  to the $N_i$ neighbors of the particle $i$ at time $t_0$. As CR quantities, we have measured the mean-squared displacement $\Delta r^2_{CR}$ and the self-part of the intermediate scattering function $F^{CR}(q,t)$ that is defined as follows
\begin{align}
    F^{CR}(q,t) = \left\langle \sum_k e^{-i \Delta \rr(t)^{CR} \cdot \mathbf{q} } \right\rangle_s \; .
\end{align}
Through $F^{CR}$ we compute the structural relaxation time $\tau_\alpha$ defined as $F^{CR}(\tau_\alpha) = e^{-1}$. To provide evidence of dynamical heterogeneity, we also computed the dynamical susceptibility $\chi_4(t)$ that measures the sample-to-sample fluctuations of $F^{CR}(q,t)$ \cite{berthier2011dynamical}. To visualize dynamical heterogeneity, we measure the map of displacement performed by each particle on a time scale of the order of the structural relaxation time $\tau_\alpha$.

\subsection*{Data availability}
The data that support the findings of this study are available from the corresponding author upon reasonable request.

\subsection*{Code availability}
The code is available from the corresponding author upon reasonable request.

\subsection*{Acknowledgments}
We thank Chiu Fan Lee and Francesco Ginelli for their comments.
M.P. acknowledges European Union's Horizon 2020 (MSCA grant agreement No 801370), the Secretary of Universities 
and Research of the Government of Catalonia (Beatriu de Pin\'os Program Grant No. BP 00088 (2018)), NextGeneration EU (CUP B63C22000730005) within the Project IR0000029 - Humanities and Cultural Heritage Italian Open Science Cloud (H2IOSC) - M4, C2, Action 3.1.1. 
D.L. acknowledges MICINN/AEI/FEDER for financial support under grant agreement RTI2018-099032-J-I00.
I.P. acknowledges MICINN, DURSI, and SNSF for financial support under
Projects No. PGC2018-098373-B-I00, No. 2017SGR-884,
and No. 200021-175719, respectively.

\subsection*{Author contributions}
M.P., D.L., and I. P. designed the research and discussed the results. M.P. performed simulations and data analysis. M.P, D.L., and I.P contributed to the writing of the manuscript.

\subsection*{Competing interests}
The authors declare no competing interests.

\bibliography{biblio}
\bibliographystyle{rsc}

\clearpage

\end{document}